\documentclass[prd,twocolumn,showpacs,preprintnumbers,tightenlines,amssymb,
               eqsecnum,superscriptaddress,floatfix]{revtex4}
\usepackage{epsfig,dcolumn,amsmath}

\begin{document}
\preprint{gr-qc/0301060}
\preprint{AEI-2003-006}

\author{Nigel T. Bishop}
\affiliation{Department of Mathematics, Applied Mathematics and
Astronomy, University of South Africa, P.O. Box 392, Pretoria
0003, South Africa}
\affiliation{Max-Planck-Institut f\"ur Gravitationsphysik,
Albert-Einstein-Institut, Am M\"uhlenberg 1, D-14476, Golm, Germany}
\author{Florian Beyer}
\affiliation{Max-Planck-Institut f\"ur Gravitationsphysik,
Albert-Einstein-Institut, Am M\"uhlenberg 1, D-14476, Golm, Germany}
\author{Michael Koppitz}
\affiliation{Max-Planck-Institut f\"ur Gravitationsphysik,
Albert-Einstein-Institut, Am M\"uhlenberg 1, D-14476, Golm, Germany}

\title{Black hole initial data from a non-conformal decomposition}


\begin{abstract}
We present an alternative approach to setting initial data in general
relativity. We do not use a conformal decomposition, but instead
express the 3-metric in terms of a given unit vector field and one
unknown scalar field. In the case of axisymmetry, we have written a
program to solve the resulting nonlinear elliptic equation. We have
obtained solutions, both numerically and from a linearized analytic
method, for a perturbation of Schwarzschild.

\end{abstract}

\pacs{04.20.Ex, 04.25.Dm, 04.70.Bw}

\maketitle

\section{Introduction}
Interferometry antennas (e.g.\ LIGO
\cite{detectors}) all over the world 
are going online to measure very fine dynamical variations of
spacetime curvature known as gravitational waves. It is hoped that
this will serve as a further evidence concerning
Einstein's theory of general relativity. Eventually it might lead to a very new
kind of observational astronomy because gravitational waves emitted by astronomical
sources reach us more or less unattenuated and give us information
when other astronomical sources block the electromagnetic spectrum.
Our understanding of gravitational waveforms emitted
by various events is limited in the strong field case, particularly with
processes involving black holes. These are important because they
are likely to be powerful and therefore detectable~\cite{cut-th}.
So for some years there has been a considerable effort to solve the
binary black hole collision problem numerically.

In order to solve Einstein's equations numerically
one needs initial 
data, i.e. the description of the state of the system
on an initial Cauchy surface. The usual approach is to
use the conformal method, in which it is assumed that the initial
data is conformally related to a reference metric. Often the reference
metric is flat because the problem is then much simpler, although 
other cases have been investigated (see below).
The literature on the subject is quite
extensive; we do not discuss it in any detail, but instead refer to a recent
review~\cite{cook}. The motivation for using conformally flat initial
data as done by e.g. Misner~\cite{misner} and Bowen and
York~\cite{bowenyork}, is mathematical
convenience rather than physics.  As
was pointed out in~\cite{garat_price} conformally
flat slices of the Kerr metric are not likely to exist. Even in the case of
non-spinning black holes it was shown~\cite{lousto_price} that
astrophysically correct initial data cannot be represented this way. 

Conventional wisdom holds that the unphysical part of an initial data
set will be radiated away from the system, within the order of the
light-crossing time. While this may well be true, the hypothesis has
not been well tested. In order to do so, initial data sets
constructed by different methods are needed. Of course such methods,
including the one presented here, will involve some form of mathematical
convenience. Provided the methods are really different, the influence
of the precise form of the intial data on the subsequent spacetime
evolution, can be investigated.

The construction of black hole initial data that is not conformally flat,
has been accomplished in recent years. The conformal method with a
non-flat reference metric in Kerr-Schild form has been used to
construct black hole data~\cite{matzner,marro1,marro2}; see also\cite{pfei}.
The Kerr-Schild ansatz which was introduced in~\cite{bishop} was a promising
ansatz that is very different to the conformal approach. However, only
the case of a
particular perturbation was solved in~\cite{bishop}, and subsequent
developments have been limited~\cite{moreno}. The motivation for the
present work was to find a variation of the ansatz of~\cite{bishop}
for which there is
a prospect of being able to find a solution in a variety of circumstances.
The ansatz presented here satisfies this condition. It leads to a
(nonlinear) elliptic equation in one unknown, and we have written a
program to solve the equation. The initial data equation has been solved
for a perturbation of Schwarzschild, using both the numerical
program as well as an analytic method that linearizes the problem.

Much use is made of computer algebra in this work, and we have used both
Maple and Mathematica so as to provide a check on our calculations.

We first present our ansatz and the constraint equations
(Sec.~\ref{s:mod}). Next, in Sec.~\ref{s:sc}, we express initial data for
the Schwarzschild geometry in terms of our ansatz. In Sec.~\ref{s:pe} we
use our ansatz to find initial data for a perturbation of
Schwarzschild, solving the problem both numerically and by linearization.
Sec.~\ref{s:co} summarizes our results and also discusses possible further
work.

\section{Modification of the Kerr-Schild Ansatz}
\label{s:mod}
We attempted to modify the Kerr-Schild ansatz so that the constraint equations would
reduce to a single elliptic equation -- because then, under many circumstances, a
solution to the problem can be expected to exist. The simplest way to do this is to
require that the extrinsic curvature be zero
\begin{equation}
K_{ij}=0
\label{eq-mom}
\end{equation}
and the momentum constraints are automatically satisfied. The Hamiltonian constraint
reduces to
\begin{equation}
{}^{(3)}R=0,
\label{eq-ham}
\end{equation}
and we used computer algebra to investigate the nature of Eq.~(\ref{eq-ham}) for
various forms of the 3-metric that involve a unit vector field and one scalar
function in analogy with \cite{bishop}. The following ansatz seems to be acceptable
\begin{equation}
\gamma_{ij}  =  \frac{d_{ij} - 2 V k_i k_j}{1 - 2V}
\label{eq-ansatz}
\end{equation}
with $d_{ij}$ the Euclidean (flat) 3-metric, $V$ a scalar field, and $k_i$ the
normalized gradient of another scalar field
\begin{equation}
k_i=\frac{\Phi_{,i}}{ \sqrt{d^{ij} \Phi_{,i} \Phi_{,j}} }.
\label{eq-k}
\end{equation}

It is useful to note a number of points about the ansatz defined by 
Eqs.~(\ref{eq-ansatz}), (\ref{eq-mom}) and (\ref{eq-k}):
\begin{itemize}
\item Under the circumstances described later, the ansatz leads to a nonlinear
elliptic equation for which a solution can be explicitly constructed. Whether the
ansatz leads to solutions in more general circumstances is, of course, unknown.
\item The assumptions~(\ref{eq-mom}) and (\ref{eq-k}), were made to simplify the
problem. To what extent
these assumptions can be relaxed, needs to be investigated. In order for the method
to be astrophysically useful, some relaxation will be necessary so as to be able to
allow for situations where there is momentum and spin.
\end{itemize}

We have evaluated the Hamiltonian constraint in the case of axisymmetry, i.e.
$x^i=(r,\theta,\varphi)$, $d_{ij}=$ diag$(1,r^2,r^2\sin^2\theta)$, and with $V$ and
$\Phi$ depending only on $r$ and $\theta)$. The reason for the restriction to
axisymmetry was problem simplification, while at the same time including non-trivial
situations such as two black holes and a perturbed Schwarzschild black hole. We used
computer algebra to find
\begin{align}
\label{eq:Ham_Constraint}
c_{20} V_{,rr}&+c_{11} V_{,r\theta}+c_{02} V_{,\theta\theta}+c_{5} \left(V_{,r}\right)^2
      +c_{4} V_{,r} V_{,\theta}\notag\\
&+c_{3} \left(V_{,\theta}\right)^2+c_2 V_{,\theta}+c_1 V_{,r}+c_0 V=0.
\end{align}
The coefficients are functions of $r$, $\theta$, $V$, $\Phi$ and higher derivatives of $\Phi$.
Some of the coefficients are very long expressions, and so here we give only the coefficients
of the principal part
\begin{align*} 
  c_{20}&=-2\frac{\left( 1 - 2\,V \right) \,
       {\Phi_{,\theta}}^2 + 
      2\,r^2\,{\Phi_{,r}}^2 }{
    \left( 1 -2\,V \right) \,
    \left( {\Phi_{,\theta}}^2 + 
      r^2\,{\Phi_{,r}}^2 \right) }\\
  c_{11}&=-4\frac{\,\left( 1 + 2\,V \right) \,
    \Phi_{,\theta}\,\Phi_{,r}}
    {\left( 1 - 2\,V \right) \,
    \left( {\Phi_{,\theta}}^2 + 
      r^2\,{\Phi_{,r}}^2 \right) }\\
  c_{02}&=-2\frac{2\,{\Phi_{,\theta}}^2 + 
    r^2\,\left( 1 - 2\,V \right) \,
     {\Phi_{,r}}^2}{r^2\,
    \left( 1 - 2\,V \right) \,
    \left( {\Phi_{,\theta}}^2 + 
      r^2\,{\Phi_{,r}}^2 \right) }  
\end{align*}
The determinant of the principal part is 
\[\Delta=c_{20}c_{02}-\frac 14 c_{11}^2=\frac{8}{r^2\,\left( 1 - 2\,V(r,\theta ) \right) },\] 
showing that the equation is elliptic provided $V<1/2$.

\section{Schwarzschild geometry}
\label{s:sc}
We now find an explicit representation of the Schwarzschild geometry that satisfies the
ansatz~(\ref{eq-ansatz}). This will be used for a number of purposes, including the
setting of boundary data in more general situations, and being the zeroth order solution
about which a perturbed Schwarzschild solution will be constructed.

Consider the Schwarzschild 3-metric in  isotropic coordinates $(\bar{r},\theta ,\varphi)$
\begin{equation}
ds^2 = \left( 1+ \frac{m}{2 \bar{r}} \right)^4 \left( d\bar{r}^2
            + \bar{r}^2 d\theta^2 + \bar{r}^2 \sin^2\theta d\varphi^2 \right)
\label{eq-sch}
\end{equation}
with the extrinsic curvature of the spacelike initial slice $K_{ij}=0$. We can obtain
the form~(\ref{eq-ansatz}) by requiring
\begin{equation}
\label{eq:coordinate_trafo}
\left( 1+ \frac{m}{2 \bar{r}} \right)^2 d\bar{r} = dr
\end{equation}
which is easily integrated to give the coordinate transformation
\begin{equation}
\label{eq:r_rbar}
r = \bar{r} + m \ln \frac{2 \bar{r}}{m} -\frac{m^2}{4 \bar{r}} +
\frac{m}{2}.
\end{equation}
where the integration constant has been chosen so that the event horizon is at $r=\bar r=m/2$.
Then in $(r,\theta ,\varphi)$ coordinates the metric satisfies ansatz~(\ref{eq-ansatz}) with
\begin{equation}
\label{eq:Schwarzschild_V}
\Phi = \frac{1}{r}, \;\;\;
V(r)= V_S(r) \equiv \frac{1}{2} \left( 1- \left( 1+\frac{m}{2 \bar{r}} \right)^{-4}
                  \frac{r^2}{\bar{r}^2} \right)
\end{equation}
On the event horizon at $r=m/2$ we have 
\begin{equation}
V=\frac{15}{32}, 
\end{equation}
and also we can make an asymptotic expansion to find
\begin{equation}
\lim_{r \rightarrow \infty} V(r)= \frac{m}{2 r} \left( 1- 2\ln \frac{2r}{m} \right).
\end{equation}
As an additional check on our calculations, we substituted Eq.~(\ref{eq:Schwarzschild_V}) into
Eq.~(\ref{eq:Ham_Constraint}), obtaining $0=0$ as expected.

\section{Perturbation of the Schwarzschild Metric}
\label{s:pe}
We now explore the case of a perturbed single Schwarzschild black hole with mass $m=1$.
We set  $\Phi=1/r$ and use the simple inner boundary condition 
\begin{equation}
\label{eq:InnerBoundary}
V\left(r=\frac 12,\theta\right)=\frac{15}{32}+\epsilon \,P_n(\cos\theta)
\end{equation}
to perturb the black hole where $P_n$ is the $n$-th Legendre polynomial. In principle,
the size of $\epsilon$ is  only limited by the fact that $V<1/2$ to preserve ellipticity
of Eq.~\eqref{eq:Ham_Constraint}, so
\[|\epsilon|<\frac 1{32}.\]
At the analytic level, the outer boundary condition is $V(r,\theta) \rightarrow V_S(r)$ as
$r \rightarrow \infty$, but computationally we set $V(r,\theta)$ to $V_S(r)$ at $r=r_{OB}$
which makes physical sense when $r_{OB}$ is large.

We solve the problem in two ways. First, we linearize, that is we ignore all terms in
Eq.~\eqref{eq:Ham_Constraint} of order $\epsilon^2$, and we are able to write the solution
as an infinite sum of eigenfunctions. Second, we develop a numerical method to solve directly the
elliptic problem Eq.~\eqref{eq:Ham_Constraint}, and use the linearized solution as an
analytic solution against which the numerical method can be validated. The reason for doing
this is that we expect, in future work, to apply the numerical method to problems for which
an approximate analytic solution cannot be found.

\subsection{Linearized Analysis}
\label{s-lin}
If the perturbation introduced in Eq.~\eqref{eq:InnerBoundary} is
small enough, i.e.\ $\epsilon\ll 1$, it is a good
approximation to neglect terms of order $\epsilon^2$ and higher. In
Sec.~\ref{sec:numerics} we consider how small $\epsilon$ should be. 

It turns out that the
ansatz 
\begin{equation}
\label{eq:Linear_Ansatz}
\Phi=1/r, \;\;\; V\left(r,\theta\right)=V_S(r)+\epsilon w_n(r) P_n(\cos\theta)
\end{equation}
separates the linearized Hamiltonian constraint and we are left with an ordinary differential
equation for $w_n(r)$
\begin{equation}
d_1 w_n^{\prime\prime}(r)+d_2 w_n^\prime(r)+d_{3(n)} w_n(r)=0
\label{eq-wn}
\end{equation}
The coefficients were found using computer algebra
\begin{align*}
d_1=&\,r^2\,{\left( -1 + 2\,V_S(r) \right) }^2\\
d_2=&-\,r\,\left( -1 + 2\,V_S(r) \right) \,
  \left( 3 - 6\,V_S(r) + 7\,r\,V_S'(r) \right)\\
d_{3(n)}=&1 - \frac{n(n+1)}2 + \left(3n(n+1)-8 \right) \,V_S(r) \\
  &- 2\,\left(3n(n+1)-10 \right) \,{V_S(r)}^2 \\
  &+ 
  4\,\left(n(n+1)-4 \right) \,{V_S(r)}^3 + 
  7\,r^2\,{V_S'(r)}^2.
\end{align*}
The fact, that the Schwarzschild solution $V_S(r)$ (Eq.~\eqref{eq:Schwarzschild_V}) is itself a
solution of the Hamiltonian constraint, has been used. We will define the functions $w_n(r)$
to be those solutions of\eqref{eq-wn} which satisfy the boundary conditions
\begin{equation}
w_n(r=\frac{1}{2})=1,\;\;\; w_n(r \rightarrow \infty) = 0,
\end{equation}
(although, in numerical calculations, it will be convenient to impose the outer
boundary condition at finite $r$).

Of course, a general solution to the perturbed Schwarzschild problem can be constructed by
summing the eigenfunctions, i.e.
\begin{equation}
V\left(r,\theta\right)=V_S(r)+\epsilon \sum^\infty_{n=1} a_n w_n(r) P_n(\cos\theta)
\end{equation}
where the $a_n$ are arbitrary constants.

\subsubsection{The York tensor}
In order to show that the 3-metric defined by Eq~\eqref{eq:Linear_Ansatz} is not conformally
flat, we need to prove that the York tensor~\cite{eisenhart,york:dof}
\begin{equation}
Y_{ijk} = R_{ij;k}-R_{ik;j} + \frac{1}{4} (R_{,j} g_{ik} - R_{,k} g_{ij}),
\end{equation}
does not vanish identically. We have used computer algebra to find $Y_{ijk}$:
if only the first eigenfunction is present (i.e., if $n=1$) then the York tensor is zero
to first order in $\epsilon$; but, for example,
\[Y_{112}= \epsilon \frac{2-n(n+1)}{4\,r^2} w_n(r) P_n'(\cos\theta ).\]
So it is proven that the eigenfunctions with $n>1$ do not represent a conformally flat geometry.

\subsection{Numerical Computation}
\label{sec:numerics}
\subsubsection{Implementation}
We want to solve Eq.~\eqref{eq:Ham_Constraint}, subject to the boundary condition defined
by Eq.~\eqref{eq:InnerBoundary} and
\[\left.V(r,\theta)\right|_{r=r_{OB}}=V_S(r)\]
with $V_S(r)$ given by Eq.~\eqref{eq:Schwarzschild_V},
with a standard numerical elliptic solver. The computations were done
using the Cactus-Computational-Toolkit \cite{cactus} and the
TAT-Jacobi elliptic solver \cite{TATJacobi} using the Jacobi method
\cite{langtangen} implementing Eq.~\eqref{eq:Ham_Constraint} and the
boundary conditions by means of second order finite differencing.

Due to the symmetry of the problem we impose the additional boundary conditions
\[\left.\frac{\partial V}{\partial n}\right|_{\theta=0,\pi}=0\]
where $n$ is normal to the $\theta=0,\pi$-surfaces.
For simplicity we implemented these
conditions at $\theta=0+\eta,\pi-\eta$ with $\eta\ll 1$ and of the
order of magnitude of the accuracy up to which Eq.~\eqref{eq:Ham_Constraint} is to be solved.
This is a common
procedure to avoid numerical problems at the singular points of the equation
($\theta=0,\pi$) -- for example, the same situation arises in the case of the Laplace
equation in spherical coordinates.

To solve non-linear elliptic PDEs using the Jacobi method the choice
of the initial guess for $V(r,\theta)$ is crucial. It turns out to be
sufficient to use Eq.~\eqref{eq:Schwarzschild_V} plus a Legendre
perturbation of a given order $n$
with linearly decreasing amplitude from the inner to the outer boundary consistent with the
boundary conditions above. To compute $V_S(r)$ for the initial guess numerically,
Eq.~\eqref{eq:r_rbar} was solved for
$\bar r$ by a numerical integration of Eq.~\eqref{eq:coordinate_trafo} which was
then substituted into Eq.~\eqref{eq:Schwarzschild_V} -- the problem being that
Eq.~\eqref{eq:r_rbar} is explicit in the wrong direction.

The convergence of TAT-Jacobi is very slow mainly due to
non-linear terms in Eq.~\eqref{eq:Ham_Constraint} which are dominant close
to the horizon. So we first run the elliptic solver with the
residual multiplied by  $(1-2V)$, which is small near the horizon.
This gives the solver the opportunity to get an accurate
solution everywhere else before, in a second run, we solve the
original equation. By means
of this technique the convergence speed was significantly increased.   

\subsubsection{Results}
We investigate the case $n=2$ with the outer boundary at
$r_{OB}=10$ and $\epsilon=0.005$. We show that three different
resolutions exhibit second order convergence, and compare the numerical
results to the linearized ones of Sec.~\ref{s-lin}.  The next table
describes the different resolutions used. With \textit{final
residual} we mean the residual of Eq.~\eqref{eq:Ham_Constraint} after
the elliptic solver has finished.

\begin{center}
\begin{tabular}{|cccccc|}\hline
   & $N_r$ & $N_{\theta}$ & $\Delta_r$ & $\Delta_{\theta}$ & final resid. of Eq.~\eqref{eq:Ham_Constraint}\\\hline
low & $97$ & $33$ & $0.098$ & $0.095$ & $4.0\cdot 10^{-7}$\\
medium & $193$ & $65$ & $0.049$ & $0.048$ & $1.0\cdot 10^{-7}$\\
high & $385$ & $129$ & $0.025$ & $0.024$ & $2.5\cdot 10^{-8}$\\\hline
\end{tabular}
\end{center}

\begin{figure}[!t]
\epsfig{file=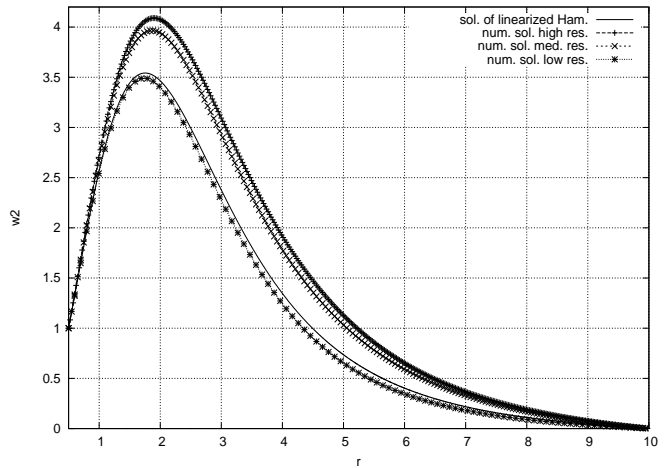,height=2.5in,width=3.5in,angle=0}
  \caption{Linearized and full numerical solutions for $n=2$, $\theta=0.59$}
  \label{fig:lin_num}
\end{figure}

\begin{figure}[t]
  \centering
\epsfig{file=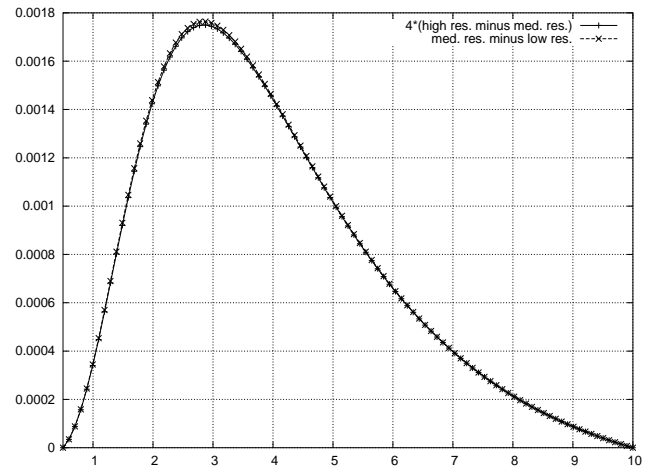,height=2.5in,width=3.5in,angle=0}
  \caption{Confirmation of second order convergence in the case $n=2$, $\theta=0.59$}
  \label{fig:convergence}
\end{figure}

Fig.~\ref{fig:lin_num} shows the full numerical results together with the linearized
one at $\theta=0.59$. In the numerical case, we plot the difference
$V(r,\theta)-V_S(r)$ and normalized to unity at $r=\frac{1}{2}$.
The graphs suggest second order convergence which is confirmed by
Fig.~\ref{fig:convergence}. It is obvious that the linearized solution is not the
limit of infinite resolution.

\begin{figure}[t]
  \centering
  \epsfig{file=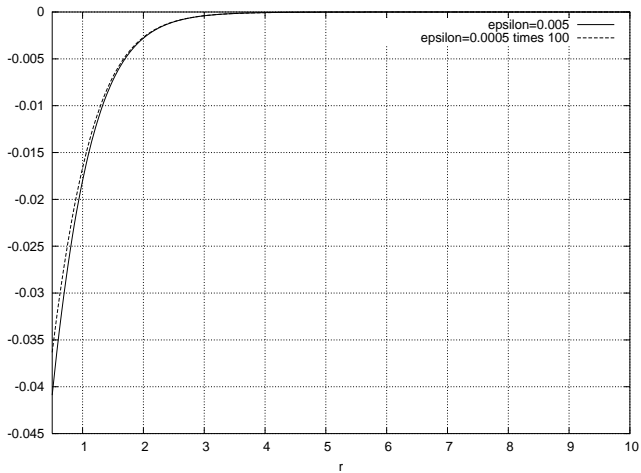,height=2.5in,width=3.5in,angle=0}
  \caption{Residuum of Eq.~\eqref{eq:Ham_Constraint} for the linearized solution}
  \label{fig:error_lin}
\end{figure}

We have thus shown that our numerics show the right convergence, but
it must still be understood why the linearized solution deviates so
much from the numerical ones. The answer is that higher order terms
get large close to the horizon. Using computer algebra, we substituted the
linearized solution which fulfills the linearized Hamiltonian constraint
Eq.~\eqref{eq-wn} up to an
error of $10^{-6}$ into Eq.~\eqref{eq:Ham_Constraint}. The residual is
shown in Fig.~\ref{fig:error_lin}, and we also show 100 $\times$ the residual
when $\epsilon$ is smaller by a factor of 10, i.e. $\epsilon=0.0005$.
Although the linearized solution is
not accurate close to the boundaries for $\epsilon=0.005$, the residual shows the
right quadratic scaling for decreasing $\epsilon$ due to second order
terms. This gives us a measure of how small $\epsilon$ should be in order to
obtain a desired accuracy.

\begin{figure}[!t]
  \centering
  \epsfig{file=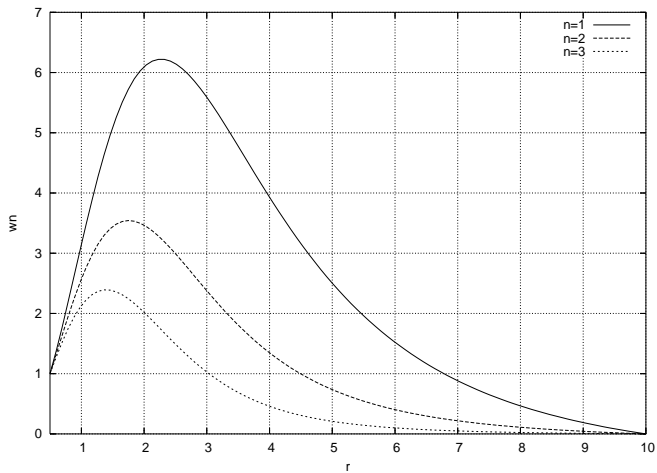,height=2.5in,width=3.5in,angle=0}
  \caption{Linearized solutions for different $n$}
  \label{fig:lin_diff_n}
\end{figure}
The computations were repeated for different positions of the
boundaries and different $n$, recovering convergence as
before. Fig.~\ref{fig:lin_diff_n} shows the linearized solutions --
i.e.\ full solutions provided $\epsilon$ is small enough -- for
different $n$.

\section{Conclusions}
\label{s:co}
The paper has presented a new method of finding initial data for the Einstein equations.
The method is not based upon a conformal decomposition, but instead a unit vector field
is chosen and then the Hamiltonian constraint reduces to a partial differential
equation in one unknown. At least in the case considered here -- zero extrinsic
curvature, axisymmetry, and with
the vector field being a normalized gradient of a scalar field -- the Hamiltonian
constraint is elliptic, and we have developed a numerical programme for solving this
equation. We have tested our numerical routine with boundary data corresponding to both the
unperturbed and perturbed Schwarzschild solutions, and the observed second order
convergence is a validation of the numerical method.

At the analytic level, we have constructed a coordinate transformation from isotropic
Schwarzschild coordinates into coordinates in which the metric satisfies our ansatz.
Further, we have investigated small perturbations of Schwarzschild, and
have shown that in this case the Hamiltonian constraint is separable so that its
solution may be written as an infinite sum of eigenfunctions. The York tensor of the
perturbed solution has been found: except for the case of the first eigenfunction, 
the York tensor is non-zero and thus the 3-metric is not conformally flat.

Two immediate items of further work are envisaged
\begin{itemize}
\item In the perturbed Schwarzschild, case it will be interesting to evolve the
initial data obtained here and compare it to an evolution from initial data obtained
from the standard conformal decomposition.
\item The two black hole problem needs to be tackled. This will be difficult technically
because at some point
between the two black holes the unit vector field must be singular, which means that
some of the coefficients in Eq.~(\ref{eq:Ham_Constraint}) will be singular. At the
analytic level this simply means that $V=0$ at the singular point, but the existence of
a singularity will complicate the numerics.
\end{itemize}
The conformal method for finding initial data has been known and refined over many years.
Our method requires further development -- and in particular a solution to the two black
hole problem -- before some form of systematic comparison between the two methods would
be worthwhile. Even so, the existence of an alternative to the conformal method may be useful
in investigating how different initial data sets, representing the same physics, affect
spacetime evolution.

\begin{acknowledgments}

We thank Sergio Dain for discussions.
This work was partially supported by the National Research Foundation, South
Africa under Grant number 2053724.

\end{acknowledgments}

\bibliography{Paper_InitialData.bib}
\end{document}